\documentclass[11pt,twoside]{article}

\usepackage{asp2006}
\usepackage{graphicx,amssym}

\newcommand{\oii}{[O{\sc ii}] }

\begin{document}
\title{The evolution of the colour-magnitude relation and of the star formation
  activity in galaxy clusters since $z\sim 0.8$}
\author{Gabriella De Lucia}
\affil{Max Planck Institut f\"ur Astrophysik, Postfach 1317, D-85748 \\
Garching bei M\"unchen, Germany}
\author{Bianca M. Poggianti}
\affil{INAF - Osservatorio Astronomico di Padova, I-35122, Italy}
\begin{abstract} 
  We present recent results on the evolution of the colour-magnitude relation
  and of the star formation activity in galaxy clusters since $z\sim 0.8$.
  Results are based on the ESO Distant Cluster Survey (EDisCS) - an ESO large
  programme aimed at the study of cluster structure and cluster galaxy
  evolution over a significant fraction of cosmic time - and are discussed in
  the framework of the current standard paradigm of structure formation.
\end{abstract}

\section{Introduction}   

Galaxy clusters are commonly considered as good laboratories for studying the
physical processes that drive galaxy evolution. This foremost role is primarily
due to the practical advantage of having many galaxies (all approximately at
the same redshift) in a relatively small region of the sky, so that efficient
observations can be carried out even with modest fields of view, and with
limited amounts of telescope time. Within our current standard paradigm for
structure formation, galaxy clusters represent, however, {\it biased}
environments originating from the highest (and rarest) peaks in the primordial
density field, whose evolution appears to be more {\it rapid} and perhaps more
{\it violent} than in regions of the Universe with average density (see
discussion in De Lucia 2007).

The instrumental capabilities achieved in recent years, and the completion of
modern large photometric and spectroscopic surveys have provided a rapidly
growing database of high--redshift clusters, and widened the mass range of the
examined systems. It is now becoming evident that the physical properties of
cluster galaxies depend not only on redshift but also on cluster mass, and that
this dependency needs to be taken into account for a correct interpretation of
observational data in terms of evolution. In this framework, the EDisCS cluster
sample represents an ideal dataset. 

EDisCS provides homogeneous photometry and spectroscopy for $20$ fields
containing galaxy clusters from $z\sim0.4$ to $z\sim 0.8$.  Cluster candidates
were selected from the Las Campanas Distant Cluster Survey \citep{Gonzalez01}
by identifying positive surface brightness fluctuations in the background sky.
The parent EDisCS sample was constructed by selecting $30$ from the highest
surface brightness candidates, and confirming the presence of an apparent
cluster and of a possible red-sequence with shallow VLT exposures in two
filters. $20$ cluster candidates were then followed up with deep optical
photometry with FORS2/VLT \citep{White_etal_2005}, near-IR photometry with
SOFI/NTT (Arag{\'o}n-Salamanca et al. in preparation), and multislit
spectroscopy with FORS2/VLT \citep{Halliday04,Milvang-Jensen08}.

\begin{figure}
\begin{center}
\hspace{-1.4cm}
\resizebox{8cm}{!}{\includegraphics{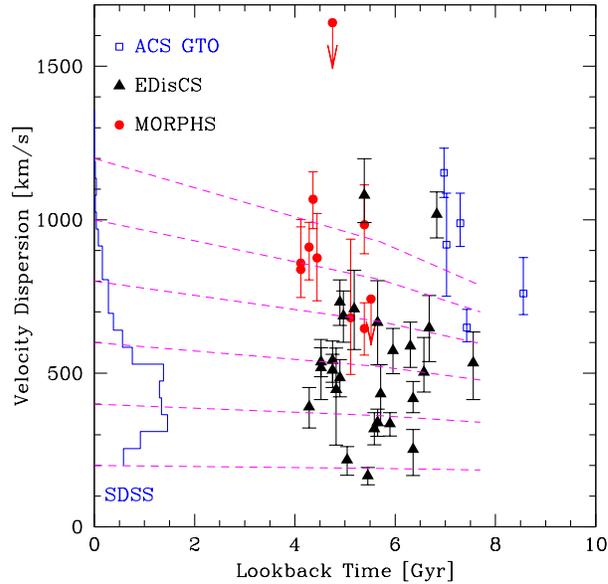}}\\%
\caption{From Milvang-Jensen et al.~(2008). Velocity dispersion as a function
  of lookback time for the EDisCS clusters (black triangles), for two other
  cluster samples at similar redshifts (MORPHS - red circles, and ACS GTO -
  blue squares), and for a local sample from the Sloan Digital Sky Survey (SDSS
  - blue histogram). The dashed lines show theoretical predictions for the
  evolution of velocity dispersions from $z=1$ to $z=0$.}
\label{fig:veldisp}
\end{center}
\end{figure}

At the time of writing, all phases of the programme have been completed, and
most of the data have been made publicly
available\footnote{http://www.mpa-garching.mpg.de/galform/ediscs/}. A first
basic characterisation of our sample is presented in White et al.~(2005) and
shows that the EDisCS sample covers a wide range of masses and structural
properties. Fig.~\ref{fig:veldisp} shows that the EDisCS clusters have velocity
dispersions that are generally lower than those of other well studied clusters
at similar redshift. They fill quite nicely the redshift gap between the MORPHS
and the ACS GTO clusters which are the only two large high-redshift cluster
samples studied in similar details. Perhaps more interestingly,
Fig.~\ref{fig:veldisp} shows that the EDisCS sample spans the appropriate range
of cluster velocity dispersions for comparison with local samples.

In this review, we summarise the results of two studies based on EDisCS data
and related to the evolution of the star formation activity
(Sec.~\ref{sec:sfr}) and of the colour-magnitude relation (Sec.~\ref{sec:cm}).

In the following, we will assume a $\Lambda$CDM cosmology: $\Omega_{\rm m} =
0.3$, $\Omega_{\Lambda}=0.7$, and $H_0 = 70\,{\rm km}\,{\rm s}^{-1}\,{\rm
  Mpc}^{-1}$. With this cosmology, $z \sim 0.8$ - the highest redshift probed
by the EDisCS sample - corresponds to more that $50$ per cent of the look-back
time to the Big Bang. All magnitudes used in the following are Vega magnitudes.

\section{The star formation activity}
\label{sec:sfr}

\begin{figure}
\resizebox{8.cm}{!}{\includegraphics[]{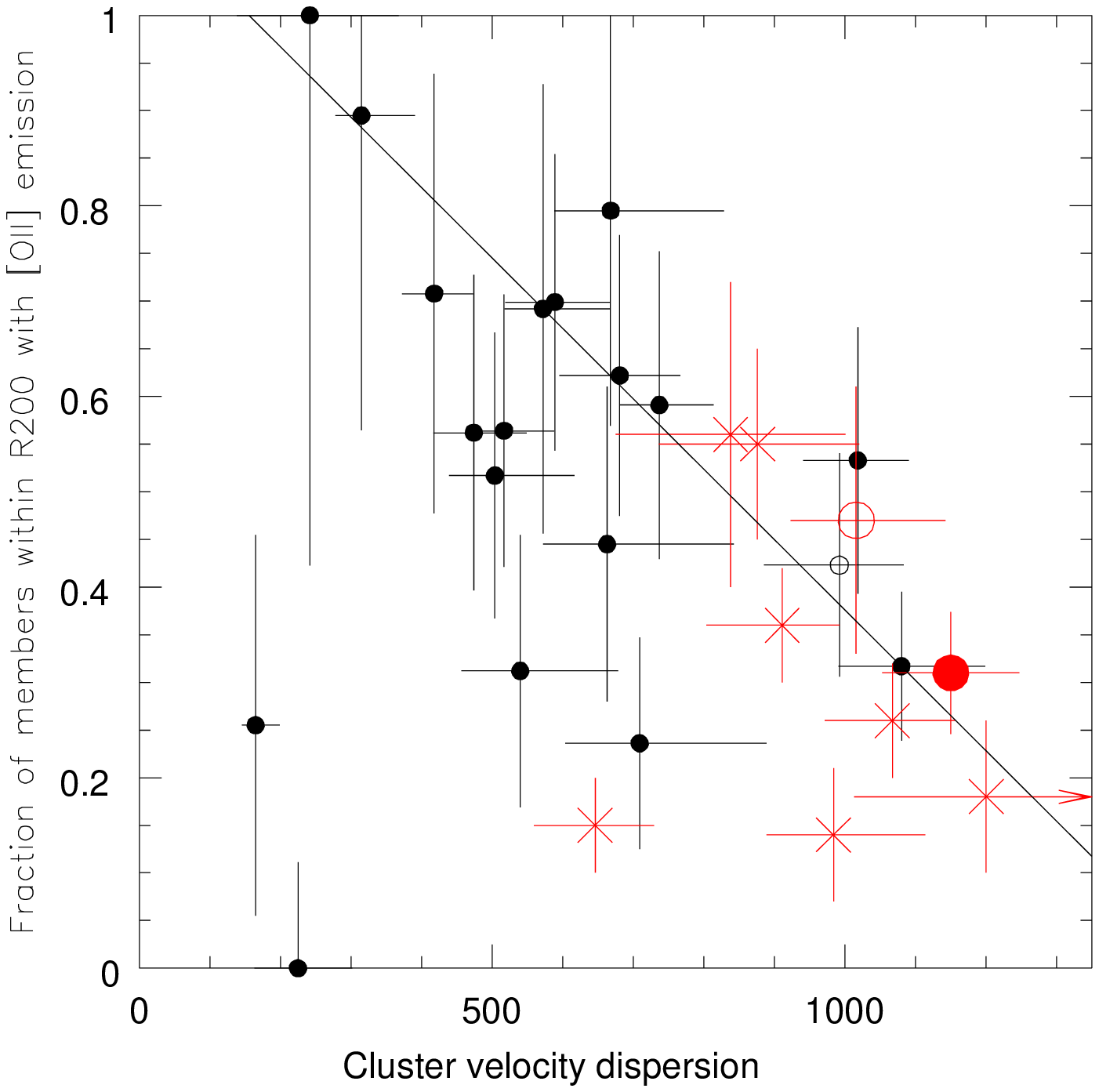}}
\hspace{-1.5cm}
\resizebox{8.cm}{!}{\includegraphics[]{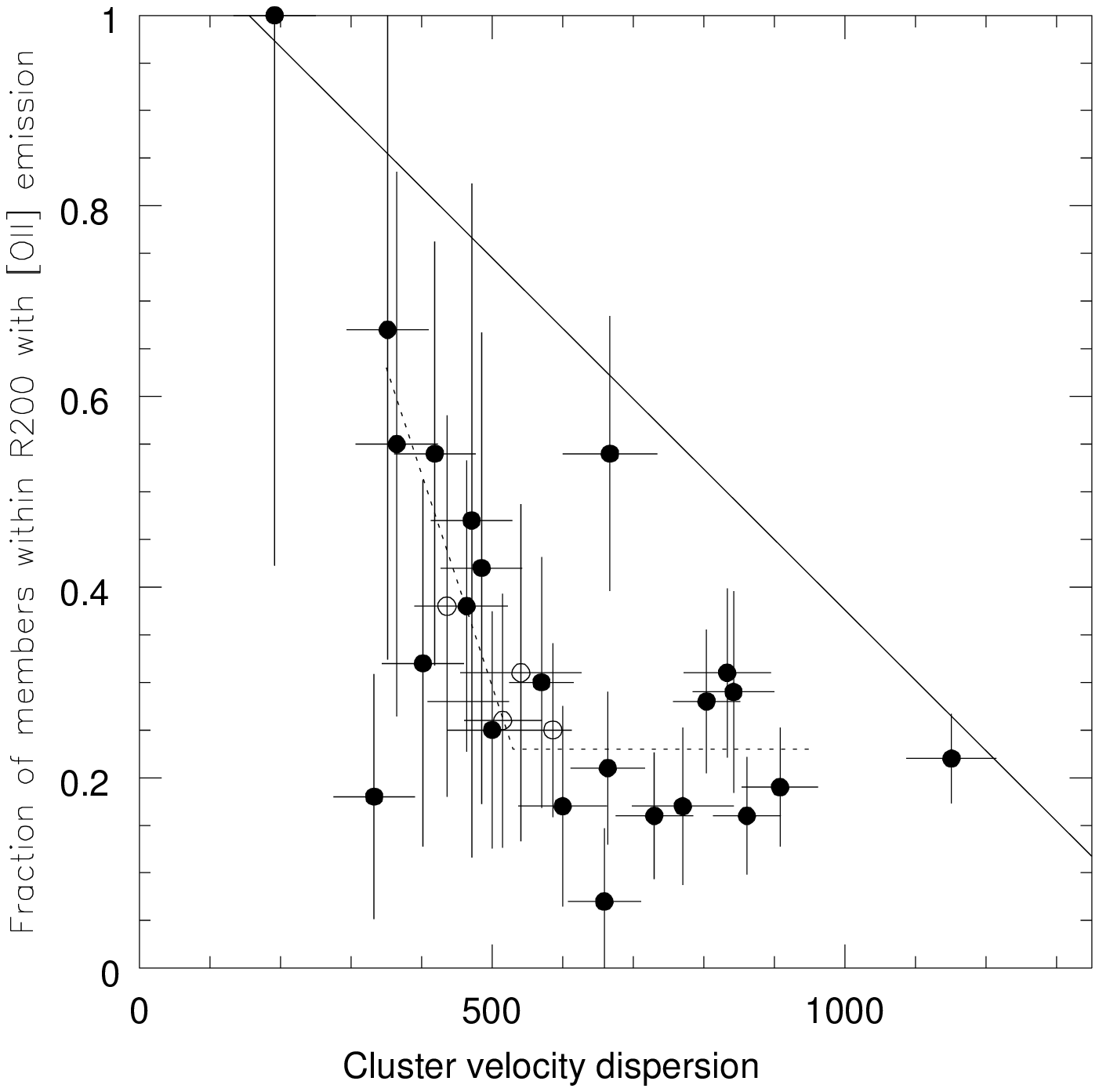}}
\caption{From Poggianti et al.~(2006). The \oii - $\sigma$ relation in clusters
  at $0.4 \le z \le 0.8$ (left panel), and for low redshift clusters from SDSS
  (right panel). On the left panel, solid black dots represent the EDisCS
  clusters, and red symbols corresponds to literature data for clusters at
  similar redshift. The solid line in both panels and the dotted lines in the
  right panel represent a by-eye description of the upper envelope of the high
  redshift data, and of the most populated regions for the low redshift data.}
\label{fig:f4sfr}
\end{figure}

Fig.~\ref{fig:f4sfr} shows the fraction of galaxies with \oii emission as a
function of the cluster velocity dispersion for clusters in the redshift range
$0.4-0.8$ in the left panel, and for a sample of low-redshift clusters from the
SDSS in the right panel. All galaxies with an EW of \oii $< -3\,\AA$ and within
the cluster virial radius ${\rm R}_{200}$ are considered (for details, see
Poggianti et al.~2006).  The error bars are large and there is a large scatter
for clusters with similar velocity dispersion. The general trend, however, is
of a broad anti-correlation for high redshift clusters, where data indicate the
presence of an `envelope' in the sense that more massive clusters have {\it at
most} a certain fraction of star forming galaxies. Interestingly, the trend is
not visible when considering only the most massive systems. For clusters at low
redshift, the fraction of star forming galaxies is approximately constant for
systems with velocity dispersion larger than $\sim 500\,{\rm km}{\rm s}^{-1}$,
and shows a very large scatter for systems with lower velocity dispersion.

\begin{figure}
  \resizebox{6.3cm}{!}{\includegraphics[bb= 62 204 232 365,clip]{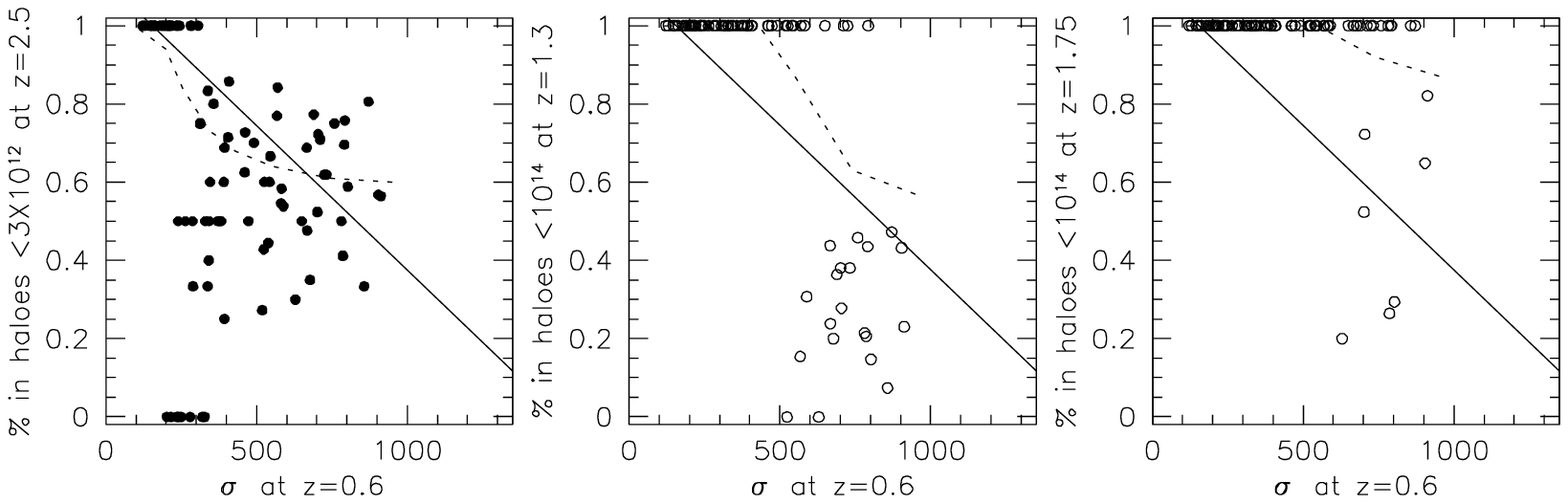}}
  \resizebox{5.9cm}{!}{\includegraphics[bb= 319 190 548 416,clip]{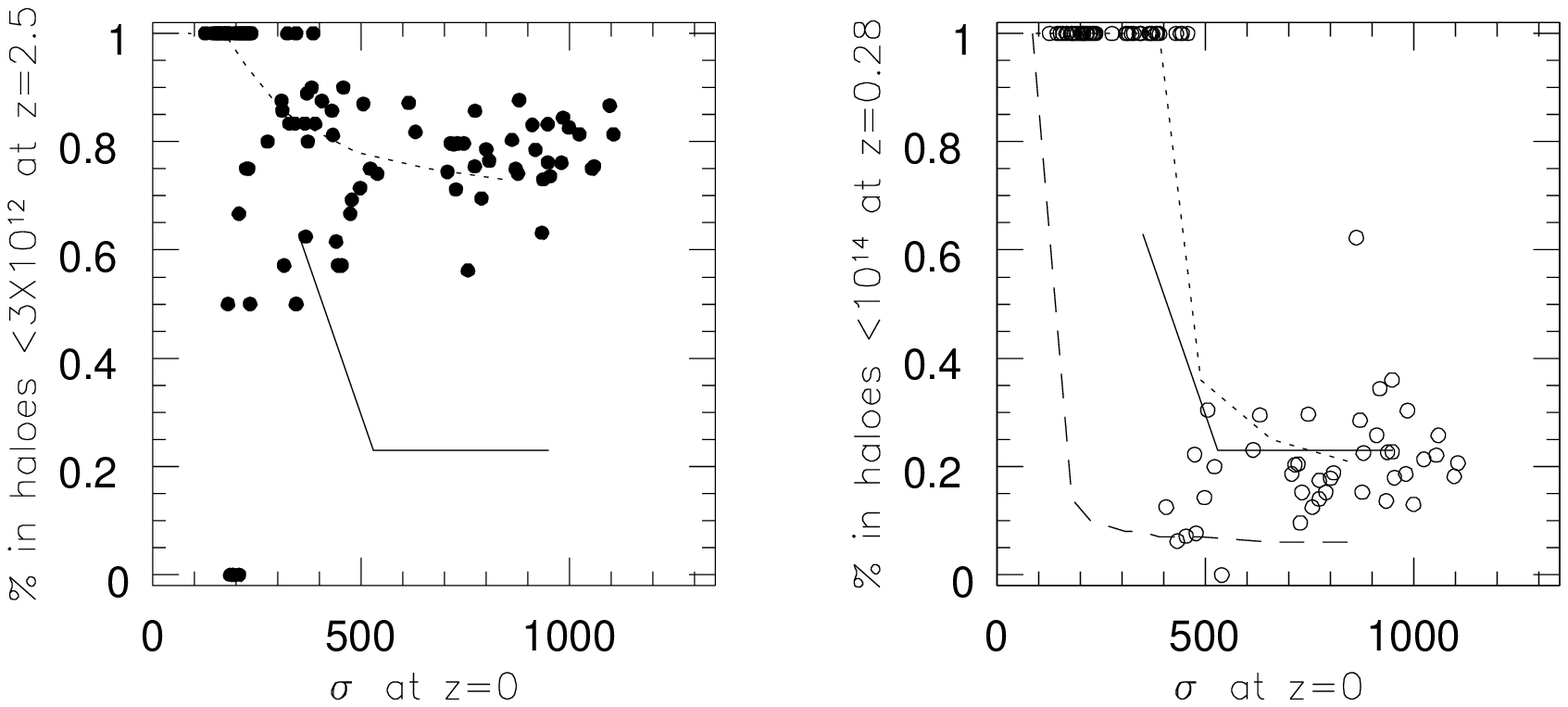}}
\caption{From Poggianti et al.~(2006). The solid lines in both panels
  correspond to the solid and dashed lines in Fig.~\ref{fig:f4sfr}. In both
  panels, symbols represent simulated haloes from the Millennium Simulation,
  while dashed lines are predictions obtained using the extended
  Press-Schechter formalism (see original paper for details).}
\label{fig:sfrev}
\end{figure}

In order to investigate a possible connection between the observed trends and
the growth history of cosmic structures, we used two different approaches: (i)
the extended Press \& Schechter formalism was used to estimate the fraction of
final mass that resides in haloes of different mass at earlier times; (ii)
semi-analytic techniques coupled to $N$-body simulations were used to estimate
the number of galaxies in haloes of different mass at higher redshifts. The
results of this analysis are shown in Fig.~\ref{fig:sfrev}. The figure shows
that the fraction of passive\footnote{Here we consider `passive' all galaxies
  with no \oii emission. The fraction of passive galaxies is therefore
  $1-f_{\oii}$ where $f_{\oii}$ is the fraction of star forming galaxies shown
  in Fig.~\ref{fig:f4sfr}} galaxies at high-z is largely determined by the
fraction of mass/galaxies that were already in haloes with mass $\gtrsim
3\times10^{12}\,{\rm M}_{\sun}$ at $z=2.5$ (left panel). The fraction of
passive galaxies in low-z clusters is in nice agreement with the fraction of
mass/galaxies in haloes with mass $\gtrsim 1\times10^{14}\,{\rm M}_{\sun}$ at
$z=0.28$ (right panel).

The redshift $z=0.28$ corresponds to a lookback time of $\sim 3$~Gyr. This
time-interval can be considered a reasonable upper limit for the time
required to extinguish star formation in newly accreted galaxies by taking into
account the time-scales associated to various physical processes that might
lead to a suppression of the star formation activity, and the
spectro-photometric timescale for the evolution of the \oii signature. The
trends shown in Fig.~\ref{fig:f4sfr} are therefore consistent with a scenario
where today cluster passive galaxies consist of two different populations:
`primordial' passive galaxies whose stars formed at $z>2.5$, and `quenched'
passive galaxies whose star formation has been suppressed at later times,
possibly by physical processes at play in dense environments.

\section{The colour-magnitude relation}
\label{sec:cm}

\begin{figure}
  \resizebox{6.6cm}{!}{\includegraphics[bb= 10 567 392 790,clip]{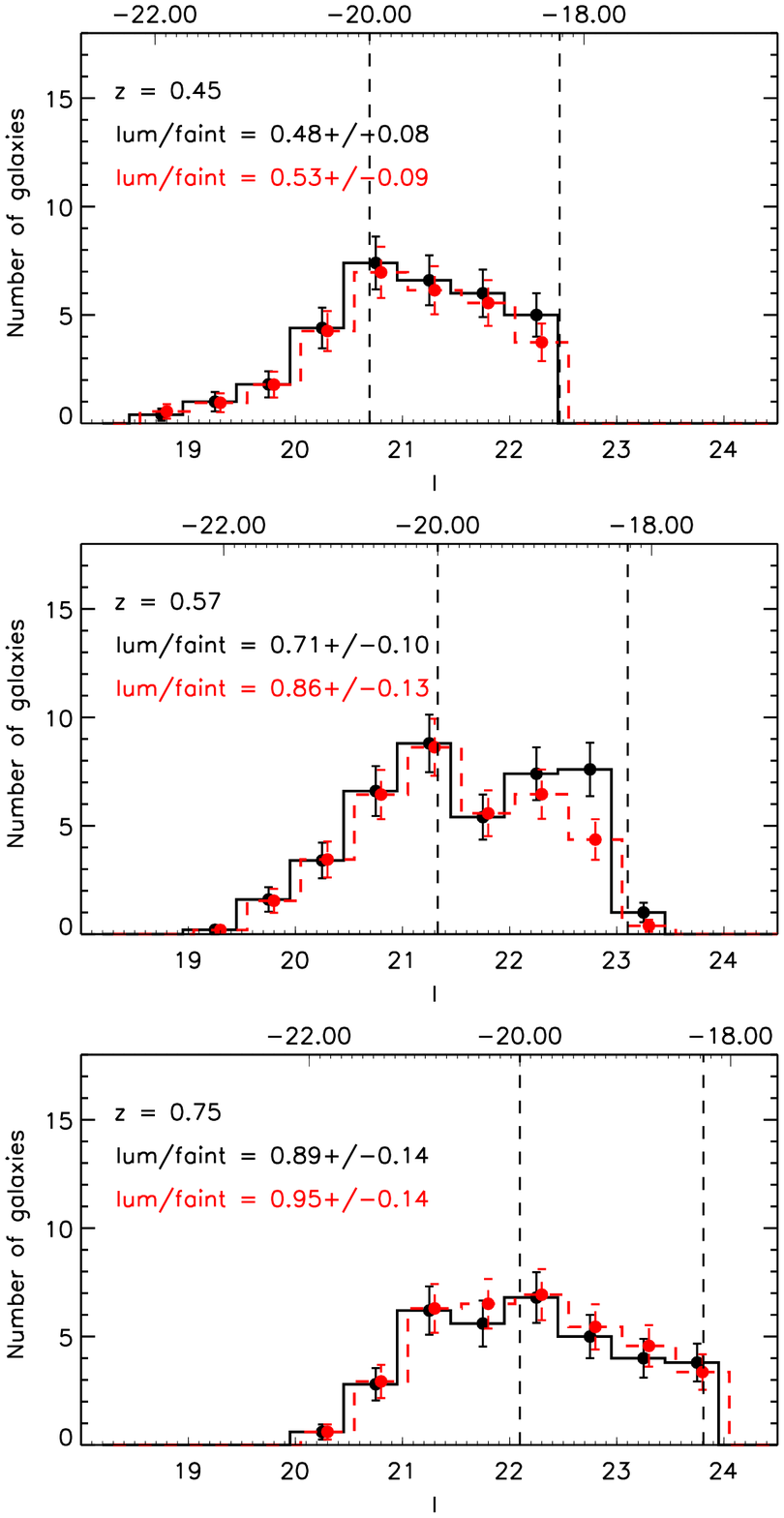}}
  \hspace{-1.cm}
  \resizebox{6.6cm}{!}{\includegraphics[bb= 10 340 392 563,clip]{delucia_fig4a.eps}}
  \resizebox{6.6cm}{!}{\includegraphics[bb= 10 114 392 336,clip]{delucia_fig4a.eps}}
  \hspace{0.5cm}
  \resizebox{5.cm}{!}{\includegraphics[]{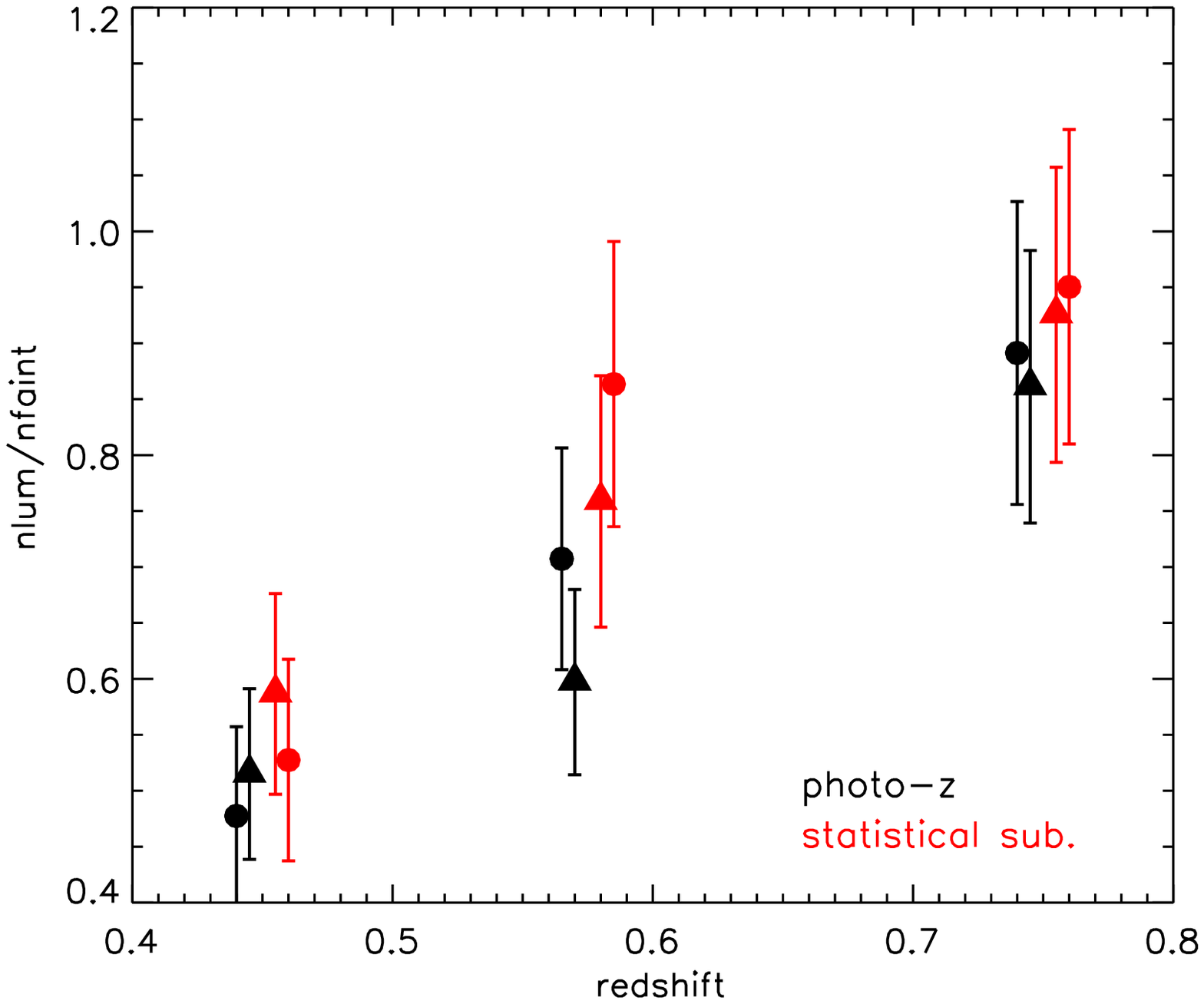}}
\caption{From De Lucia et al.~(2007). The first three panels show the number of
  red-sequence galaxies as a function of magnitude. The bottom-right panel
  shows the evolution of the luminous-to-faint ratio as a function of redshift
  (see text). Different colours correspond to different criteria for cluster
  membership (red for a method based on statistical subtraction, and black for
  a photometric redshift selection). Different symbols in the bottom right
  panels correspond to different choices for the area used for the analysis
  (see original paper for details).}
\label{fig:fig4cm}
\end{figure}

The first three panels of Fig.~\ref{fig:fig4cm} show the number of
`red-sequence galaxies' as a function of their I-band magnitude. For the
histograms shown in Fig.~\ref{fig:fig4cm}, we have used all cluster members
within 0.3~mag from the best fit colour-magnitude relation, determined using
spectroscopically confirmed members with absorption line spectra. Clusters have
been combined in three redshift bins correcting colours and magnitudes to the
central redshift of the corresponding bin. Corrections are based on a single
burst model with formation redshift $z_f = 3$ and a metallicity-dependent
normalisation\footnote{As explained in the original papers, the relation
between the metallicity and luminosity has been calibrated so as to reproduce
the metallicity-luminosity relation observed for the Coma cluster.}. We have
shown that this simple model provides a good fit to the observed red-sequence
over all the redshift range sampled by the EDisCS clusters, as well as for the
Coma cluster and for a local cluster sample from the SDSS.

Black and red histograms in Fig.~\ref{fig:fig4cm} correspond to different
criteria for cluster membership. The scale on the top of each panel shows the
rest-frame V-band magnitude corresponding to the observed I-band magnitude
after correcting for passive evolution between the redshift of the bin and
$z=0$ (for details, see De Lucia et al.~2004 and De Lucia et al.~2007). By
defining as `luminous' all galaxies brighter than $M_{\rm V} = -20$, and as
`faint' all galaxies fainter than this limit and brighter than $M_{\rm V} =
-18.2$, we find a clear trend for a decreasing luminous-to-faint ratio as a
function of redshift. The trend is confirmed independently of the method used
to select cluster members and of the area used for the analysis (bottom-right
panel of Fig.~\ref{fig:fig4cm}).

\begin{figure*}
\begin{center}
\hspace{-1.4cm}
\resizebox{10cm}{!}{\includegraphics{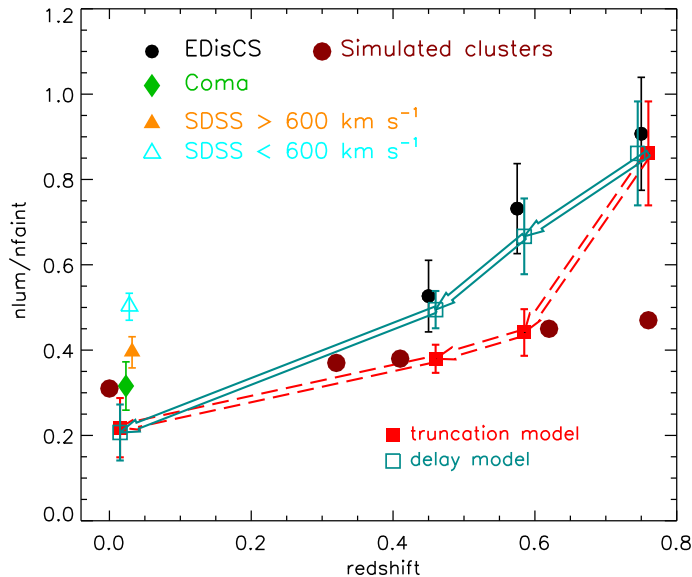}}\\%
\caption{Adapted from De Lucia et al.~(2007). Luminous--to--faint ratio as a
  function of redshift.  Black filled circles show the EDisCS data.  The green
  diamond shows the value measured for the Coma cluster.  The triangles show
  results for clusters selected from the SDSS with velocity dispersion larger
  (orange) and smaller (cyan) than $600\,{\rm km}\,{\rm s}^{-1}$.  Arrows
  indicate the evolution of the luminous--to--faint ratio obtained using the
  models described in the text. Red filled circles show predictions from the
  $N$-body simulations coupled with semi-analytic techniques.}
\label{fig:fig9cm}
\end{center}
\end{figure*}

The trend of Fig.~\ref{fig:fig4cm} can be reproduced by simple evolution of the
combined blue and red galaxies that populate the colour-magnitude diagram of
our high-redshift clusters. This is shown explicitly in Fig.~\ref{fig:fig9cm}.
The average behaviour of the EDisCS data is shown with black circles, while the
blue and red symbols connected by arrows represent the evolution predicted by
simple models obtained evolving all galaxies at earlier redshifts. Red arrows
correspond to a model where the star formation histories of the galaxies under
consideration are truncated at the redshift of the observation of each cluster,
while blue arrows correspond to a model where the star formation histories are
suppressed 1 Gyr later than the redshift of observation (see original paper for
details). The red filled circles in Fig.~\ref{fig:fig9cm} show predictions from
the semi-analytic model described in De Lucia \& Blaizot (2007).

Fig.~\ref{fig:fig9cm} shows that both the `truncation' and the `delayed' model
predict an amount of evolution from the highest redshift considered to $z=0$
that is in good agreement with that measured using the highest redshift EDisCS
clusters and the Coma cluster (green diamond). In the truncation model,
galaxies move onto the red-sequence very rapidly and the predicted
luminous-to-faint ratio lies below the measured value in the intermediate
redshift bins. In the delayed model, galaxies stay blue longer and the
predicted evolution is closer to the observed trend at all redshifts sampled by
the EDisCS clusters. The local sample of SDSS clusters used in our analysis
lies above the value measured for the Coma cluster, with luminous-to-faint
ratios compatible with those measured for the EDisCS clusters at $z\sim0.4$.
The zero-point of the colour-magnitude relation predicted by our semi-analytic
model shows the same level of evolution measured from the data, and the
luminous-to-faint ratio measured for simulated clusters at $z=0$ is in very
nice agreement with that measured for Coma. The predicted evolution as a
function of redshift is, however, weaker than observed. This indicates that in
these models the transition from blue to red occurs on a time-scale which is
shorter than observed (see also Wang et al. 2007, and discussion in De Lucia
2007).

\section{Discussion and Conclusions}

In this review, we have summarised the results of recent work studying the
evolution of the star formation activity and of the colour-magnitude relation
in galaxy clusters since $z\sim0.8$.  Our dataset is based on systems with a
wide range of velocity dispersion from the EDisCS sample, which we have
complemented with other clusters at similar redshift from the literature, and
with a local cluster sample from the SDSS. The complementarity between the
EDisCS sample and other well studied systems at similar redshift has revealed
an important element to better quantify evolution as a function of redshift
{\it and} of cluster mass\footnote{In the studies summarised here, we have used
the cluster velocity dispersion as a proxy for the mass of the system}.

Using the presence of the \oii line in emission as a signature of ongoing star
formation, we have demonstrated that the fraction of star forming galaxies
varies as a function cluster velocity dispersion, and that this dependency
changes as a function of redshift. Interestingly, our results can be explained
by linking the `passive' systems with the mass/number of objects that have
experienced different environments at earlier times. The model proposed is
quite simple and it does not provide useful insight on the physical process(es)
responsible for the suppression of the star formation in cluster galaxies. It
is, however, suggestive that a simple model based on structure formation is
able to explain quantitatively the evolution measured from our data. In fact,
structure formation is an important - and often neglected - element to be taken
into account when interpreting observational data in terms of evolution. 

Our study on the evolution of the colour-magnitude relation has demonstrated
that there is significant evolution in the luminosity distribution of
red-sequence galaxies. Compared to clusters in the local Universe, the EDisCS
clusters exhibit a significant {\it deficit} of faint red galaxies. Our results
clearly indicate that the red--sequence population of high--redshift clusters
does not contain all progenitors of nearby red--sequence cluster galaxies.  A
significant fraction of these must have moved onto the red--sequence below
$z\sim 0.8$. We have shown that the evolution in the relative distribution of
`luminous' and `faint' red-sequence galaxies can be explained by evolution of
the blue bright galaxies that populate the colour-magnitude diagram of high
redshift clusters. The model used is extremely simplified (e.g. it assumes a
single redshift of formation, it neglects the accretion of new galaxies), and
results should be interpreted with caution. A semi-analytic model coupled with
high-resolution $N$-body simulations predicts a much smaller evolution than
observed, and we have argued that this is due to a too rapid decline of the
star formation of galaxies accreted onto larger systems.

Future studies comparing in detail results like those presented here with
predictions from semi-analytic models coupled to high-resolution simulations,
will provide important constraints on galaxy formation models, and on the
physical processes governing evolution of galaxies in clusters.

\acknowledgements 
We acknowledge contributions from all EDisCS team. GDL thanks the organisers of
the meeting for their wonderful hospitality.

\end{document}